\documentclass[prd,aps,showpacs, superscriptaddress,10pt]{revtex4-1}
\usepackage[utf8]{inputenc}
\usepackage{amsmath}
\usepackage{amssymb}
\usepackage{epsfig}
\usepackage{bbold}

\newcommand{\e}{\mbox{e}}

\begin{document}

\title{Comment on ``Localization of 5D Elko Spinors on Minkowski Branes" }

\author{I. C. Jardim}  
\email{jardim@fisica.ufc.br} \affiliation{Departamento de F\'isica,
  Universidade Federal do Cear\'a, Caixa Postal 6030, Campus do Pici,
  60455-760 Fortaleza, Cear\'a, Brazil}

\author{G. Alencar}
\email{geova@fisica.ufc.br} \affiliation{Departamento de F\'isica,
 Universidade Federal do Cear\'a, Caixa Postal 6030, Campus do Pici,
 60455-760 Fortaleza, Cear\'a, Brazil}
 
 \author{R. R. Landim}  
\email{renan@fisica.ufc.br} \affiliation{Departamento de F\'isica,
  Universidade Federal do Cear\'a, Caixa Postal 6030, Campus do Pici,
  60455-760 Fortaleza, Cear\'a, Brazil}

 \author{R. N. Costa Filho} 
\email{rai@fisica.ufc.br} \affiliation{Departamento de F\'isica,
 Universidade Federal do Cear\'a, Caixa Postal 6030, Campus do Pici,
  60455-760 Fortaleza, Cear\'a, Brazil}

 \begin{abstract}
 We show that the statement in Yu-Xiao Liu, Xiang-Nan Zhou, Ke Yang and Feng-Wei Chen  [Phys. Rev. D 86 , 064012 (2012)]  that the zero mode of ELKO spinor is localized in some thin brane scenarios is not correct. This reopens the problem of localization of ELKO spinors.
 \end{abstract}

\maketitle

\section{A short review of the problem}

In Ref. \cite{Liu:2011nb} the authors investigated the possibility of zero mode localization of ELKO spinors in RS like scenarios. The model considered involves $\delta$ like branes and the respective generalization to smooth scenarios. The action used in the above-cited article for ELKO spinor is
\begin{equation}\label{action5}
S = \int d^{5}x\sqrt{-g}\left[-\frac{1}{4}\left(D_{M}\lambda D^{M}\bar{\lambda} + D_{M}\bar{\lambda} D^{M}\lambda\right) -\eta F(z)\bar{\lambda}\lambda\right] ,
\end{equation}
where $\eta$ is a coupling constant, $F(z)$ is a scalar function of conformal extra dimension $z$ and $D_{M}$ is the covariant derivative defined as
\begin{equation}
 D_{M}\lambda = \partial_{M}\lambda +\Omega_{M}\lambda.
\end{equation}
As show in \cite{Liu:2011nb} the nonvanishing components of spin connection are
\begin{equation}
 \Omega_{\mu} = \frac{1}{2}A'(z)\gamma_{\mu}\gamma_{5},
\end{equation}
where primes denote derivative with respect to the argument, $A(z)$ is the conformal warp factor $g_{MN} =\e^{2A(z)}\eta_{MN}$ and $\eta_{MN} = \mbox{Diag.}(-1,1,1,1,1)$.  
Taking the variation of action in respect to $\bar{\lambda}$ we obtain the equation of motion 
\begin{equation}
 D_{M}\left[\sqrt{-g}D^{M}\lambda\right] -2\eta\sqrt{-g}F(z)\lambda = 0,
\end{equation}
using the metric and the non vanishing components of spin connection, we can write the above equation in the form 
\begin{eqnarray}\label{lambda}
 &&\eta^{\mu\nu}\partial_{\mu}\partial_{\nu}\lambda -A'(z)\gamma_{5}\eta^{\mu\nu}\gamma_{\mu}\partial_{\nu}\lambda  -A'^{2}\lambda + \e^{-3A}(\e^{3A}\lambda')' -2\eta\e^{2A}F(z)\lambda = 0.
\end{eqnarray}
Due the term $A'(z)\gamma_{5}\eta^{\mu\nu}\gamma_{\mu}\partial_{\nu}\lambda $ the authors of \cite{Liu:2011nb} proposed a decomposition of ELKO field as $\lambda = \lambda_{+} +\lambda _{-}$ with
\begin{equation}
 \lambda_{\pm} = \e^{-3A/2}\sum_{n}\left[\alpha_{n}(z)\varsigma_{\pm}^{n}(x) +\beta_{n}(z)\tau_{\pm}^{n}(x)\right],
\end{equation}
 where $\varsigma_{\pm}^{n}(x)$ and $\tau_{\pm}^{n}(x)$ are two independent $4D$ ELKO field satisfying the Klein-Gordon equations $\Box\tau_{\pm}^{n}(x) = m_{n}^{2}\tau_{\pm}^{n}(x)$, $\Box\varsigma_{\pm}^{n}(x) = m_{n}^{2}\varsigma_{\pm}^{n}(x)$ and 
 the relations $\gamma^{5}\tau_{\pm} = \mp\varsigma_{\pm}$, $\gamma^{5}\varsigma_{\pm} = \pm\tau_{\pm}$. Using this decomposition in eq. (\ref{lambda}) and after some manipulations it is found that $\alpha_n=\beta_n$ and the problem is reduced to the following equation
 \begin{equation}\label{alphaF}
\alpha_{n}''(z)-\left(\frac{13A'^{2}}{4} +\frac{3A''}{2} -m_{n}^{2}+im_{n}A'(z)+2\eta\e^{2A}F(z)\right)\alpha_{n}(z)=0,
\end{equation}
with the normalization condition
\begin{equation}
 \int \alpha_{n}^{*}\alpha_{m}dz = \delta_{mn}.
\end{equation}
The eq. (\ref{alphaF}) is the general equation of localization coefficients $\alpha_{n}$. In following sections we will comment on the nonlocability for some cases considered in Ref. \cite{Liu:2011nb}.

\section{On the Localization of Free 5D Massless ELKO Spinors}
In this section we will comment on the zero mode localization of ELKO in thin branes for the cases with and without interaction. 

\subsection{The localization of the zero mode of a 5D free ELKO spinor}
As explained in \cite{Liu:2011nb} the problem is reduced to find solutions to the Schrödinger like equation
\begin{equation}
[-\partial^{2}_z+V_0(z)]\alpha_0=0
\end{equation}
with potential
$$
V_0=\frac{3}{2}A''+\frac{13}{4}A'^2
$$
and the orthonormality condition
$$
\int |\alpha_0|^2dz=1.
$$

In the thin brane scenario we have $A(z)=-\ln(k|z|+1)$ and we get the potential

$$
V_0=\frac{19k^2}{4(1+k|z|)^2}-\frac{3k\delta (z)}{1+k|z|}.
$$
The general solution is given by
\begin{eqnarray}
\alpha_{0}(z)=C_{1}(k|z|+1)^{\frac{1}{2}+\sqrt{5}}+C_{2}(k|z|+1)^{\frac{1}{2}-\sqrt{5}}{,} \label{FreeRS}
\end{eqnarray}
where $C_{1}$, $C_{2}$ are integral parameters. At this point the authors in \cite{Liu:2011nb} considered the condition that for $z\to \pm\infty$ the solution must be convergent fixing $C_1=0$. However they missed the boundary condition at $z=0$. At this point the zero mode equation gives us
$$
\alpha_o '|_+-\alpha_o '|_-=-3k\alpha_0(0),
$$
leading to the condition
$$
C_2=-\frac{\sqrt5+2}{\sqrt5-2}C_1.
$$
Therefore, if we try to fix $C_1=0$ to get a localized solution we get that it is trivial. On the other hand if we impose the boundary condition in $z=0$ our solution is not localized. Therefore this situation is in complete agreement with the fact that the smooth version used in \cite{Liu:2011nb} do not gives a localized zero mode. This is expected since in the limiting case one model should reduce to the other. For an interesting discussion of boundary conditions at $z=0$ for smooth version of RS see \cite{Jardim:2014vba}. 

\subsection{The localization of the zero mode of a 5D interacting ELKO spinor}
At this point we must comment on the results of the authors of \cite{Liu:2011nb} about localization of ELKO spinor in interaction models. The only change in the model is through the interacting term where the zero mode equation now is given by
\begin{eqnarray}
[-\partial_{z}^{2}+V_{0}(z)]\alpha_{0}(z)=0, \label{Schrodinger equation 2},
\end{eqnarray}
where
\begin{eqnarray}
V_{0}(z)=\frac{3}{2}A''+\frac{13}{4}{A'}^{2}+2\eta \text{e}^{2A}F(\phi). \label{effective potential Vz 2}.
\end{eqnarray}

In the thin brane case the authors of above-cited article choose an interacting term given by $\eta F(\phi)=M^2_{elko}$. Where $V_{0}$, in eq. (\ref{effective potential Vz 2}), is given by
\begin{equation}
V_{0}=\frac{(19+8\epsilon)k^2}{4(1+k|z|)^2}-\frac{3k\delta(z)}{1+k|z|},
\end{equation}
with $\epsilon=M_{elko}^2/k^2$. The authors considered the following as a bound solution to the zero mode equation (\ref{Schrodinger equation 2}):
\begin{equation}
\alpha_{0}(z)=\sqrt{(-1+\sqrt{5+2\epsilon})k}~(1+k|z|)^{\frac{1}{2}-\sqrt{5+2\epsilon}}.
\end{equation}

In the above solution it is required that $\epsilon>-2$. So the authors in \cite{Liu:2011nb} consider that for any $M_{elko}^2\ge0$, the ELKO zero mode could be localized on the RS$\mathrm{II}$ brane. However, they forget about the condition at $z=0$ again. The condition now is that
\begin{equation}\label{z0}
\alpha_o '|_+-\alpha_o '|_-=-3k\alpha_0(0) 
\end{equation}
which give us the condition $\epsilon=-1/2$, i.e., a tachyonic ELKO spinor in 5D with a fine tuned mass $M_{elko}^2 = -k^{2}/2$. Therefore, we found that it is impossible to localize the zero mode also in this case with $M_{elko}^2\ge0$. In fact, the general solution for the above potential, just as in the case for the non interacting model is given by
\begin{equation}
\alpha_{0}(z)=C_1(1+k|z|)^{\frac{1}{2}-\sqrt{5+2\epsilon}}+C_2(1+k|z|)^{\frac{1}{2}+\sqrt{5+2\epsilon}}.
\end{equation}

The condition (\ref{z0}) give us
$$
C_1(2-\sqrt{5+2\epsilon})+C_2(2+\sqrt{5+2\epsilon})=0,
$$
and we arrive at the same problem as in the previous case. If we choose a solution that satisfies the condition at $z=0$ it is not localized and if we choose a solution that is localized the it do not satisfies our boundary condition. Again, we find that this is is agreement with the results of \cite{Liu:2011nb} for the smooth case, where the zero mode is not localized.

\section {Conclusion}
It has been shown here that the solutions for localization of the zero mode of the ELKO spinor, in thin brane do not satisfies the boundary condition at $z=0$. The problem seems to be very similar to that of the gauge fields. Apparently the effective potential has to be consistently modified to change the boundary condition at infinity and at the origin. Possible solutions to this problems are the use of delta function interaction or geometrical couplings as proposed by the present authors in \cite{Jardim:2014vba,Alencar:2014moa, Alencar:2014fga}.  \section*{Acknowledgments}

We acknowledge the financial support provided by Funda\c c\~ao Cearense de Apoio ao Desenvolvimento Cient\'\i fico e Tecnol\'ogico (FUNCAP), the Conselho Nacional de 
Desenvolvimento Cient\'\i fico e Tecnol\'ogico (CNPq) and FUNCAP/CNPq/PRONEX.

\end{document}